\documentclass[submitting]{nst}

\usepackage{subfigure,dcolumn}
\usepackage[T2A,T1]{fontenc}
\usepackage{mhchem}
\usepackage{multirow}

\usepackage{soul}
\soulregister\cite7
\soulregister\ce7

\usepackage{hyperref}
\usepackage{soul}

\DeclareSIUnit{\eVee}{\eV\mathrm{ee}}
\DeclareSIUnit{\keVee}{\keV\mathrm{ee}}

\begin{document}
	\title{Identification of the anomalous fast bulk events in a \textit{p}-type point-contact germanium detector}
	\thanks{This work was supported by the National Key Research and Development Program of China (No. 2017YFA0402203), the National Natural Science Foundation of China (No. 11975162), and the SPARK project of the research and innovation program of Sichuan University (No. 2018SCUH0051).}
	
	\author{Ren-Ming-Jie Li}
	\affiliation{College of Physics, Sichuan University, Chengdu 610065, China}
	
	\author{Shu-Kui Liu}
	\email[Corresponding author, ]{Shukui Liu: liusk@scu.edu.cn}
	\affiliation{College of Physics, Sichuan University, Chengdu 610065, China}
	
	\author{Shin-Ted Lin}
	\email[Corresponding author, ]{ShinTed Lin: stlin@scu.edu.cn}
	\affiliation{College of Physics, Sichuan University, Chengdu 610065, China}
	
	\author{Li-Tao Yang}
	\affiliation{Key Laboratory of Particle and Radiation Imaging (Ministry of Education) and Department of Engineering Physics, Tsinghua University, Beijing 100084, China}
	
	\author{Qian Yue}
	\affiliation{Key Laboratory of Particle and Radiation Imaging (Ministry of Education) and Department of Engineering Physics, Tsinghua University, Beijing 100084, China}
	
	\author{Chang-Hao Fang}
	\affiliation{College of Physics, Sichuan University, Chengdu 610065, China}
	
	\author{Hai-Tao Jia}
	\affiliation{College of Physics, Sichuan University, Chengdu 610065, China}
	
	\author{Xi Jiang}
	\affiliation{College of Physics, Sichuan University, Chengdu 610065, China}
	
	\author{Qian-Yun Li}
	\affiliation{College of Physics, Sichuan University, Chengdu 610065, China}
	
	\author{Yu Liu}
	\affiliation{College of Physics, Sichuan University, Chengdu 610065, China}
	
	\author{Yu-Lu Yan}
	\affiliation{College of Physics, Sichuan University, Chengdu 610065, China}
	
	\author{Kang-Kang Zhao}
	\affiliation{College of Physics, Sichuan University, Chengdu 610065, China}
	
	\author{Lei Zhang}
	\affiliation{College of Physics, Sichuan University, Chengdu 610065, China}
	
	\author{Chang-Jian Tang}
	\affiliation{College of Physics, Sichuan University, Chengdu 610065, China}
	
	\author{Hao-Yang Xing}
	\affiliation{College of Physics, Sichuan University, Chengdu 610065, China}
	
	\author{Jing-Jun Zhu}
	\affiliation{College of Physics, Sichuan University, Chengdu 610065, China}

	\begin{abstract}
		The ultralow detection threshold, ultralow intrinsic background, and excellent energy resolution of \textit{p}-type point-contact germanium detector are important for rare-event searches, in particular for the detection of direct dark matter interactions, coherent elastic neutrino-nucleus scattering, and neutrinoless double beta decay. Anomalous bulk events with an extremely fast rise time are observed in the CDEX-1B detector. We report a method of extracting fast bulk events from bulk events using a pulse shape simulation and reconstructed source experiment signature. Calibration data and the distribution of X-rays generated by intrinsic radioactivity verified that the fast bulk experienced a single hit near the passivation layer. The performance of this germanium detector indicates that it is capable of single-hit bulk spatial resolution and thus provides a background removal technique.
	\end{abstract}
	
	
	\maketitle
	
	\section{Introduction}
	
	Compelling evidence from astrophysics and cosmology observation suggests $\sim$\SI{26.8}{\%} of the energy density of the universe consists of dark matter \cite{planckcollaborationPlanck2015Results2016,adrianUniverseHighDefBaby2013,bergstromDarkMatterEvidence2012}. Its nature and properties remain unknown. The \textit{p}-type point-contact germanium (\textit{p}PCGe) detector exhibits the characteristics of ultralow noise, low energy thresholds, and excellent energy resolution \cite{lukeLowCapacitanceLarge1989,barbeauLargemassUltralowNoise2007,aalsethResultsSearchLightMass2011,MethodsForObtainingCha2019}. These features uniquely support direct searches for dark matter in the form of weakly interacting massive particles (WIMPs) as well as the rare-event experiments. These detectors have been used extensively in rare-event experiments, such as the search for dark matter and studies of neutrino physics \cite{somaCharacterizationPerformanceGermanium2016,Zhi2017Characterization,yanyulu2020Study,XiaopengZhou109902,WUHuifang453,OnProportionalScintillatioN2020}. The China Dark Matter Experiment (CDEX) pursues the study of light WIMPs and related dark matter models \cite{qianDetectionWIMPsUsing2004,maResultsDirectDark2020} as well as neutrinoless double beta decay in \ce{^{76}Ge}, with the goal of establishing a ton-scale germanium detector array at the China Jinping Underground Laboratory \cite{cdexcollaborationIntroductionCDEXExperiment2013,kangStatusProspectsDeep2010,chengChinaJinpingUnderground2017,zhaoSearchLowmassWIMPs2016,liuLimitsLightWIMPs2014,EvaluationOfCosmogenicActivation2020}.
	
	Surface event (SE) performance of a \textit{p}PCGe detector has been studied, where SEs are identified as events within the upper $\sim$ \SI{1}{\mm} of the N$^+$ electrode \cite{maStudyInactiveLayer2017}. The signature, a slow rising pulse, is attributed to incomplete charge collection caused by the weak drift field and electron-hole recombination \cite{aguayoCharacteristicsSignalsOriginating2013}. Detail algorithms for differentiating surface and bulk events (BEs) are described in \cite{liDifferentiationBulkSurface2014,yangBulkSurfaceEvent2018}.
	
	An energy threshold of \SI{160}{\eVee} (electron equivalent energy), as determined by calibration with known cosmogenic X-ray peaks, was achieved using a kilogram-scale single-module \textit{p}PCGe in the CDEX-1B experiment \cite{yangLimitsLightWIMPs2018}. The bulk distribution, however, has an abnormal structure, with an extremely fast rising pulse compared to BEs. In this work, we report an investigation of the characteristics of these anomalous fast bulk events (FBEs) and the separation of BEs from FBEs inside the CDEX-1B detector through pulse shape simulation. Removing FBEs is expected to significantly reduce the CDEX background, and thus improve the sensitivity of dark matter searches.
	
	\section{Anomalous fast bulk events}
	
	\begin{figure}[htb]
		\centering
		\includegraphics[width=0.9\linewidth]{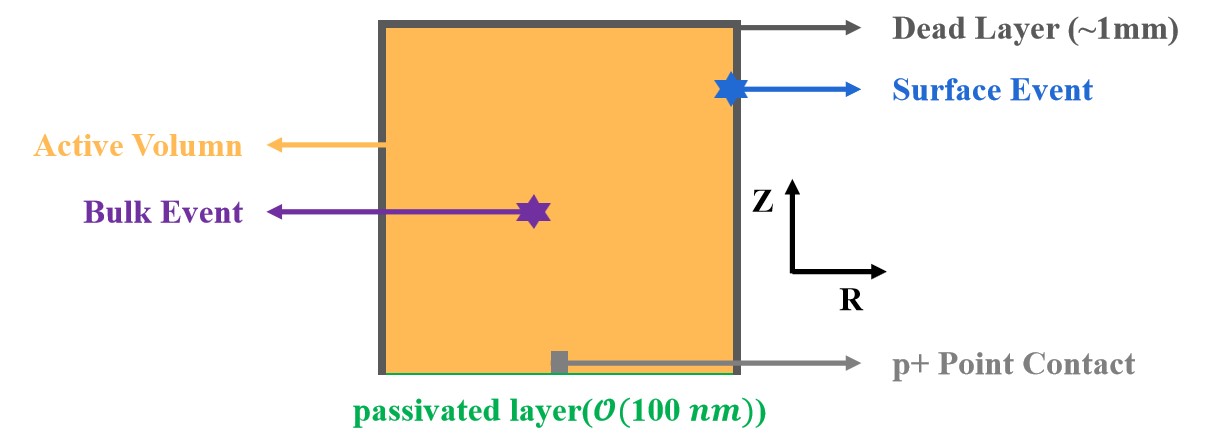}
		\caption{Schematic diagram of a typical \textit{p}PCGe detector}
		\label{fig:1-Figure}
	\end{figure}
	
	\begin{figure*}[htb]
		\centering
		\includegraphics[width=0.9\linewidth]{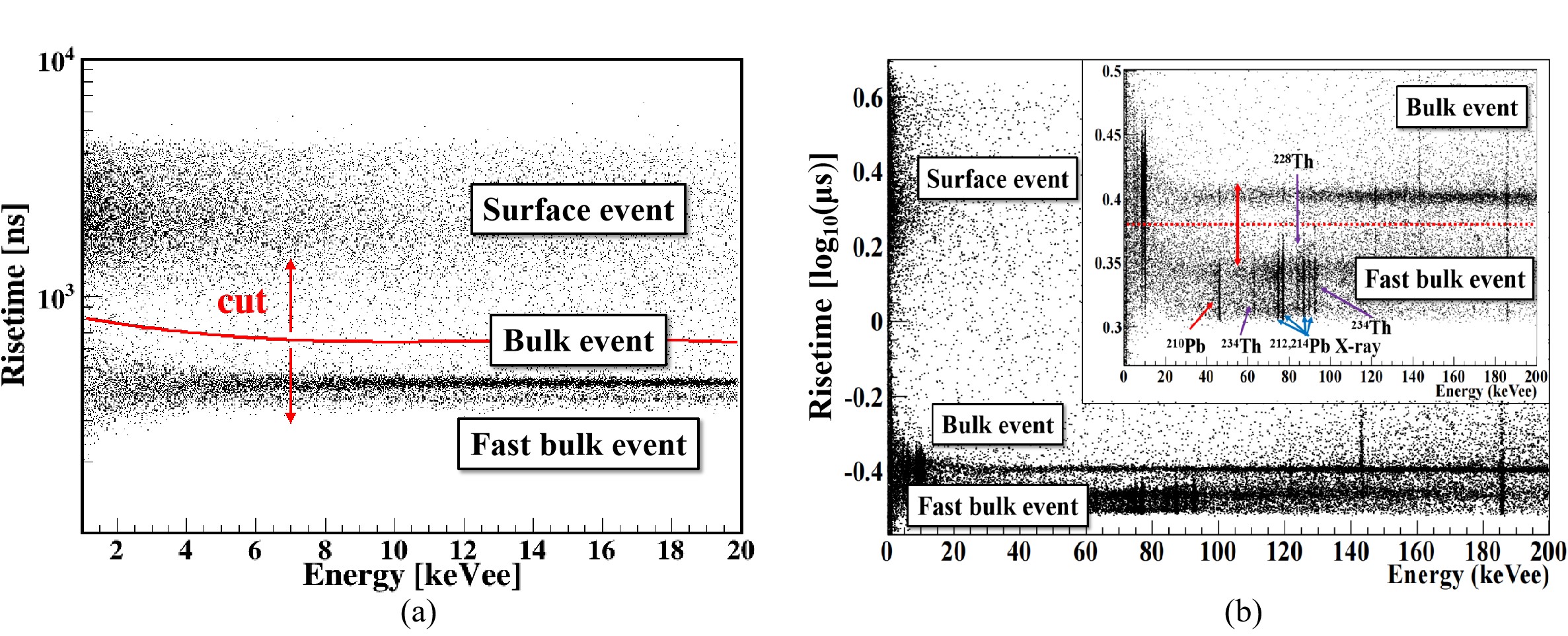}
		\caption{Scatter plots of the rise time versus energy in (a) low- and (b) high-energy regions, from CDEX-1B detector data}
		\label{fig:2-Figure}
	\end{figure*}
	
	\begin{figure*}[htb]
		\centering
		\includegraphics[width=0.9\linewidth]{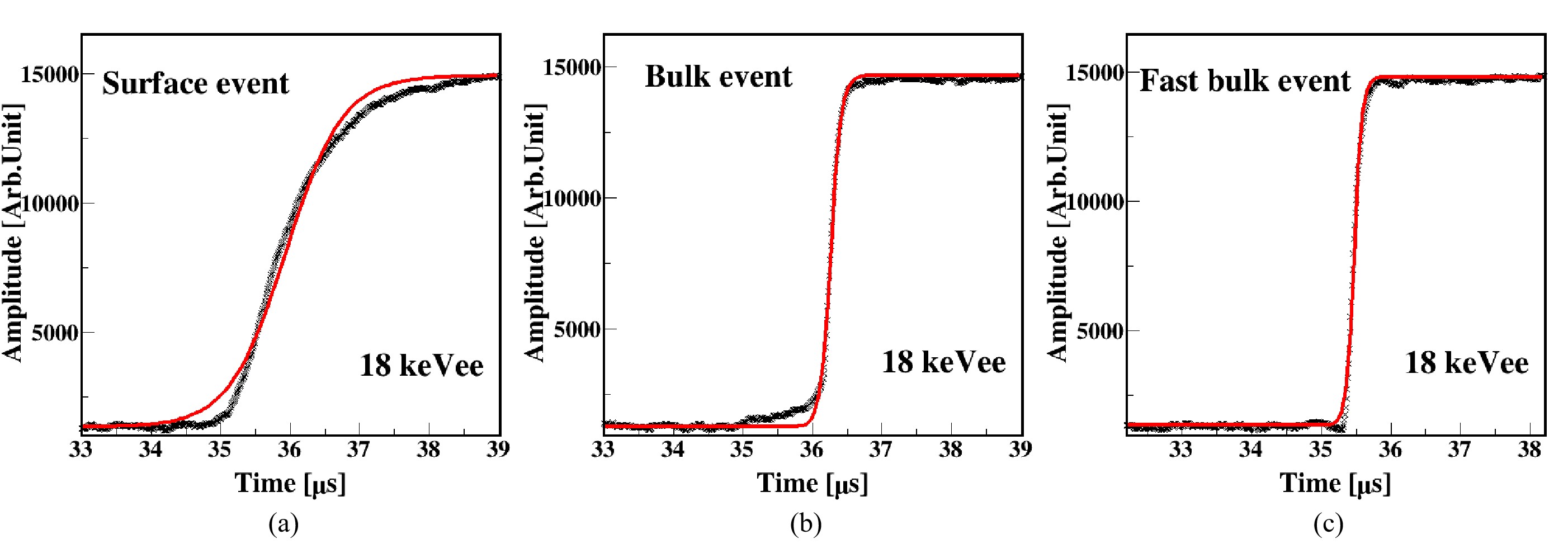}
		\caption{Typical pulses of (a) SE, (b) BE, and (c) FBE inside \textit{p}PCGe detector, with best fit of the hyperbolic tangent function (red line)}
		\label{fig:3-Figure}
	\end{figure*}
	
	\begin{figure*}[htb]
		\centering
		\includegraphics[width=0.90\linewidth]{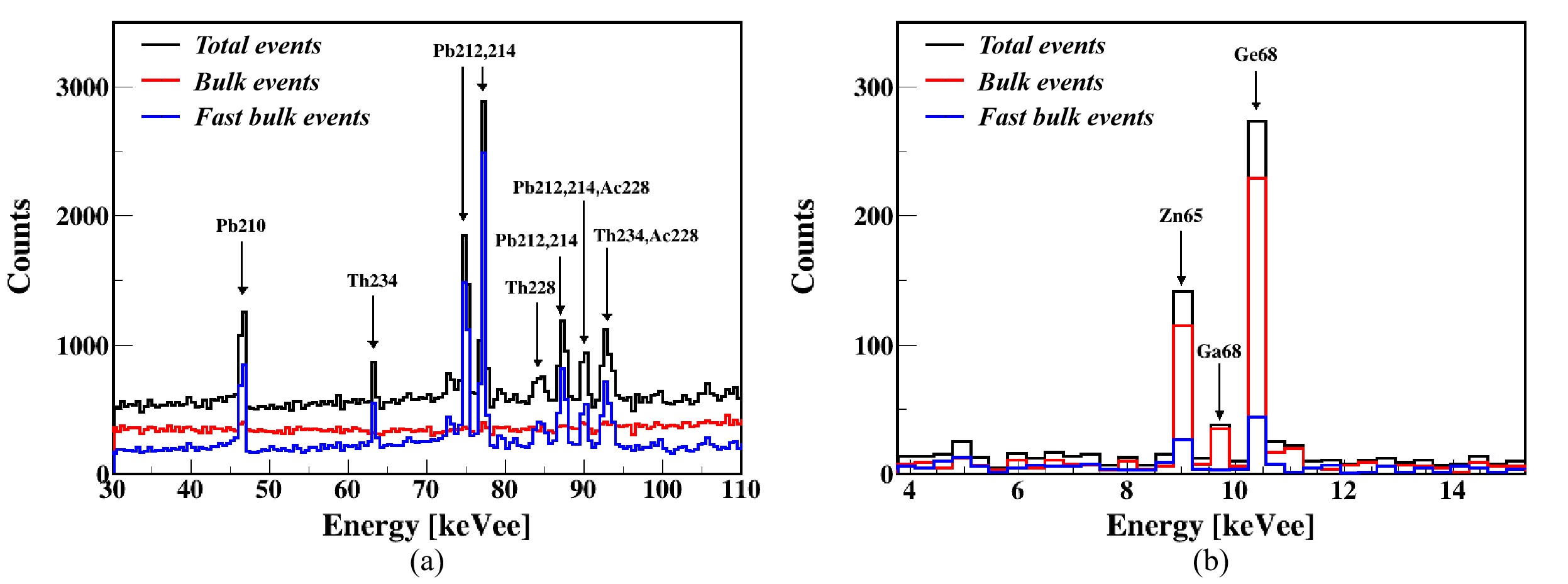}
		\caption{Background energy spectrum including BEs and FBEs in (a) relatively high-energy (\sisetup{range-phrase = --, range-units = single} \SIrange{30}{110}{\keVee}) region and (b) low-energy (\SIrange{4}{15}{\keVee}) region. FBEs are indicated by red dotted line in the inset of Fig. \ref{fig:2-Figure}(b)}
		\label{fig:4-Figure}
	\end{figure*}
	
	\begin{figure*}[htbp]
		\centering
		\includegraphics[width=0.90\linewidth]{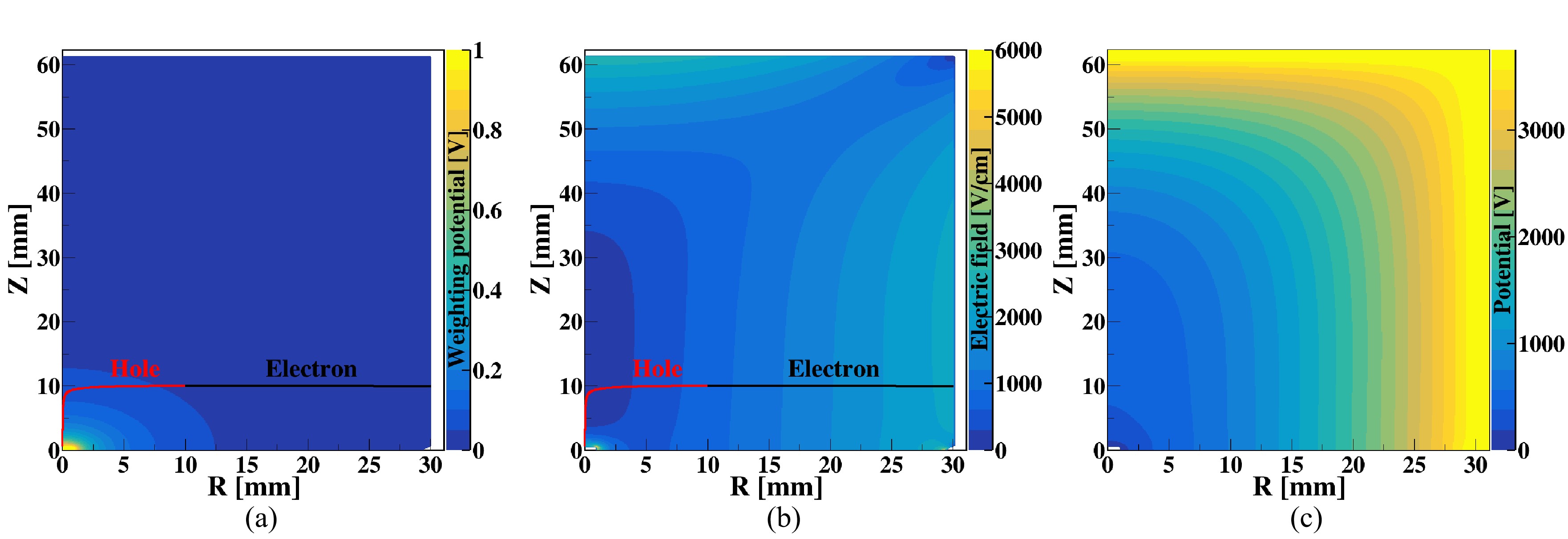}
		\caption{(a), (b) and (c) are weighting potential ($W\!P$), actual electric field, and potential of the CDEX-1B detector in the $(r,z)$ plane calculated by SAGE, where the P$^+$ point contact is at $(r = \SI{0}{\mm}, z = \SI{0}{\mm}$). The red and black trajectories in (a) and (b) represent the drift paths of holes and electrons, respectively, induced by a particle at $(r = \SI{10}{\mm}, z = \SI{10}{\mm})$}
		\label{fig:5-Figure}
	\end{figure*}
	
	Fig. \ref{fig:1-Figure} displays a schematic diagram of a typical \textit{p}PCGe detector. The \textit{p}PCGe detector has a dead layer, which is a lithium-diffused N$^+$ layer of millimeter-scale thickness \cite{liLimitsSpinIndependentCouplings2013}. In addition, there is a passivation layer of sputtered silicon oxide at the surface of the P$^+$ point contact \cite{lvevaluation2021}. It is typically several hundred nanometers thick, and alpha or beta particles can easily penetrate it and deposit energy in the detector \cite{edzardsSurfaceCharacterizationPtype2021}.
	
	Fig. \ref{fig:2-Figure}(a) and (b) shows scatter plots of rise time (time interval between \SI{5} and \SI{95}{\%} of the pulse amplitude) versus energy in the low- and high-energy regions, from data obtained using the CDEX-1B detector. The three bands, from top to bottom, indicate SEs, BEs, and FBEs, and the typical pulses of these three types of events are shown in Fig. \ref{fig:3-Figure}.
	
	The origin of the FBEs is revealed by comparing the distribution of the photopeaks in Fig. \ref{fig:2-Figure}(b) and Fig. \ref{fig:4-Figure}(a). Most of the events from all the peaks in the CDEX-1B experiment at energies of \sisetup{range-phrase = --, range-units = single}\SIrange{70}{95}{\keVee}, which result from $\gamma$ rays or X-rays from the progeny of \ce{^{238}U} and \ce{^{232}Th} \cite{DeterminationOfGross2020}, are FBEs. The CDEX-10 experiment also revealed the same behavior \cite{cdexcollaborationPerformancesPrototypePointcontact2019}. The alpha particles, beta particles, and low-energy photon produced by nearby electronics and structural materials are more likely to enter the detector through the $\mathcal{O}(100\,nm)$ thin passivation layer and contribute to the low-energy background. Because these events deposit energy close to the passivation layer, the hole carries, which make the largest contribution to the signal, are collected rapidly by the P$^+$ electrode; thus, these events have a shorter rise time than BEs. This behavior is the reason FBEs are observed, which is proved by pulse simulations, as described in the next section.
	
	The \SI{8.98}{\keVee} and \SI{10.37}{\keVee} KX peaks due to the internal cosmogenic isotopes \ce{^{65}Zn} and \ce{^{68}Ge} are presented in Fig. \ref{fig:4-Figure}(b). The ratios of FBEs to all bulk events are \sisetup{separate-uncertainty}\SI{19.4(52)}{\%} and \SI{16.2(30)}{\%}, respectively. Because these two isotopes are uniformly distributed inside the crystal, these ratios indicate the volume ratio of the FBE region to the volume of the entire crystal.
	
	Here we focus on the differentiation BEs from FBEs at energies above \SI{1.0}{\keVee} where SEs are well separated by the red line shown in Fig. \ref{fig:2-Figure}(a). Owing to the effects of electronic noise, BEs and FBEs can be roughly distinguished above \SI{10}{\keVee} and exhibit serious cross-contaminations below \SI{6}{\keVee}.
	
	\begin{figure}[htbp]
		\centering
		\includegraphics[width=0.9\linewidth]{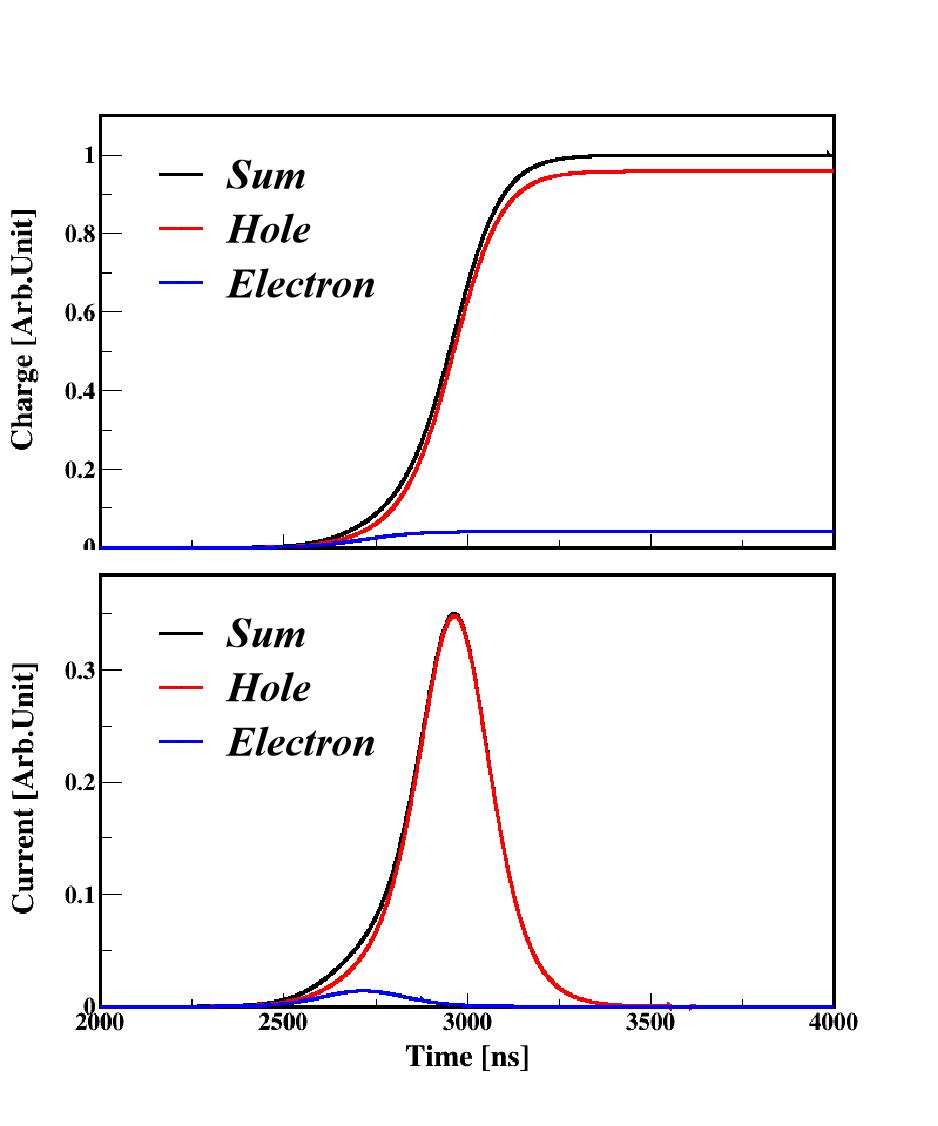}
		\caption{Simulated pulse shape with electronic response at $(r = \SI{10}{\mm}, z = \SI{10}{\mm})$ in the CDEX-1B detector, where the red and blue lines represent the hole and electron components of the signal, respectively}
		\label{fig:6-Figure}
	\end{figure}
	
	\begin{figure}[htbp]
		\centering
		\includegraphics[width=0.9\linewidth]{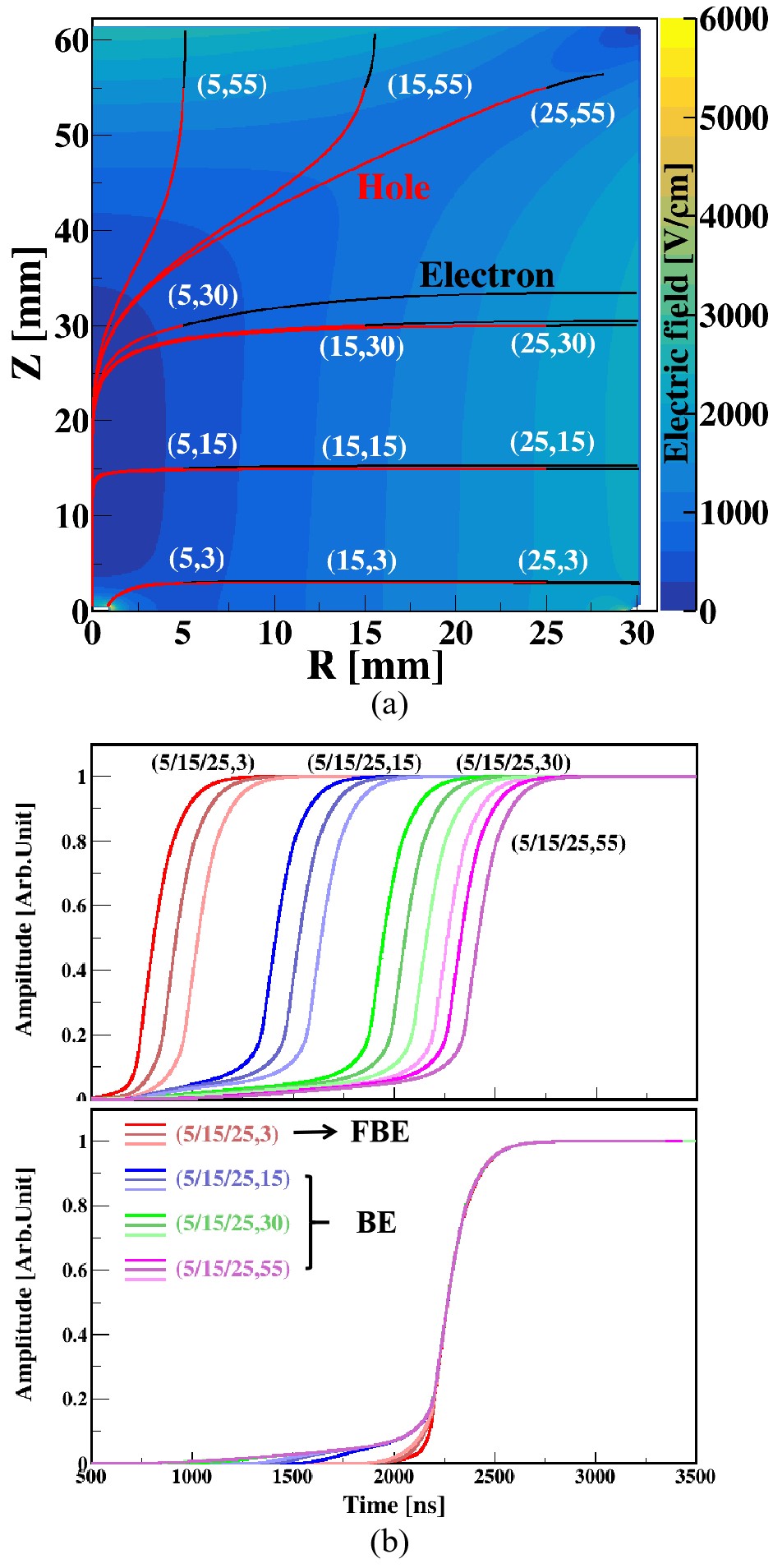}
		\caption{(a) Drift trajectories of electron-hole pair at different locations inside the detector and (b) corresponding signals with electronic response. In the bottom panel, points corresponding to half of the amplitude value of each signal are superposed}
		\label{fig:7-Figure}
	\end{figure}
	
	\section{The signal formation in \textit{p}PCGe and the pulse shape simulation}
	
	The interaction between an energetic charged particle and a germanium detector creates a crowd of charge carriers (electron-hole pairs), the number of which is proportional to the deposited energy of the particles in the detector. Then, under an electric field, these mobile charge carriers immediately drift toward their respective electrode in opposite directions. The instantaneous charge and current they induced on the contacts can be described using the Shockley-Ramo theorem \cite{shockleyCurrentsConductorsInduced1938,ramoCurrentsInducedElectron1939}. For a cluster of charge $q (q > 0)$, the time-dependent induced charge $Q(t)$ and the current $I(t)$ are expressed as
	\begin{gather}
		Q(t) = q[W\!P(\boldsymbol{r_h}(t)) - W\!P(\boldsymbol{r_e}(t))], \label{Eq1} \\
		I(t) = q[\boldsymbol{v_h}(\boldsymbol{r_h}(t)) \cdot \boldsymbol{W\!E}(\boldsymbol{r_h}(t)) - \boldsymbol{v_e}(\boldsymbol{r_e}(t)) \cdot \boldsymbol{W\!E}(\boldsymbol{r_e}(t))], \label{Eq2}
	\end{gather}
	where $W\!P(\boldsymbol{r}(t))$ and $\boldsymbol{W\!E}(\boldsymbol{r}(t))$ represent the weighting potential and weighting field, respectively, at the hole (electron) position $\boldsymbol{r_h}(t)$($\boldsymbol{r_e}(t)$). In addition, $\boldsymbol{v_h}(\boldsymbol{r_h}(t))$($\boldsymbol{v_e}(\boldsymbol{r_e}(t))$) represents the drift velocity of holes (electron) at $\boldsymbol{r_h}(t)$($\boldsymbol{r_e}(t)$). Here, the weighting potential is calculated assuming unity voltage on the P$^+$ electrode and zero voltage on all other electrodes, without considering the space charge distribution \cite{comellatoChargecarrierCollectiveMotion2021}. Therefore, the distribution of $W\!P(\boldsymbol{r}(t))$ depends only on the detector geometry and is independent of the bias voltages applied to the electrodes and the space charge distribution \cite{heReviewShockleyRamo2001}. However, the time evolution of the drift velocity and positions of charge carriers are affected by characteristics of the electric field inside the detector \cite{giovanettiPTYPEPOINTCONTACT}. Moreover, the motion of charge carriers (the charge cloud) due to thermal diffusion and Coulomb self-repulsion affects the drift paths and thus the time development of the signal \cite{mertensCharacterizationHighPurity2019}.
	
	The package SAGE \cite{sheSAGEMonteCarlo2021}, which is a Geant4-based simulation framework for the CDEX with germanium detectors, was used for the pulse shape simulation. The calculated weighting potential and the actual electric field and potential distributions of the CDEX-1B detector are shown in Fig. \ref{fig:5-Figure}, where the P$^+$ point contact is located at $(r = \SI{0}{\mm}, z = \SI{0}{\mm}$). The red and black trajectories in Fig. \ref{fig:5-Figure}(a) and (b) represent the drift paths of holes and electrons, respectively, in response to a particle at $r = \SI{10}{\mm}, z = \SI{10}{\mm}$. According to Eq. \eqref{Eq1}, at time $t = 0$, because the electrons and holes are at the same position and have the same weighting potential, $Q(t = 0) = 0$. Then, as holes drift toward the P$^+$ contact, $W\!P(\boldsymbol{r_h}(t))$ increases, whereas electrons drift toward the N$^+$ electrode, and $W\!P(\boldsymbol{r_e}(t))$ decreases. If the electrons reach the N$^+$ electrode first, $W\!P(\boldsymbol{r_e}(t)) = 0$; then $Q(t)$ is induced only by hole drift. When both electrons and holes are completely collected by the electrodes, $W\!P(\boldsymbol{r_h}(t)) = 1$, $Q(t)$ reaches the maximum value, as shown in Fig. \ref{fig:6-Figure}.
	
	The components of the charge signal, that is, the change in charge, contributed by holes [$Q_h(t)$] and electrons [$Q_e(t)$] can be expressed as
	
	\begin{gather}
		Q_h(t) = q[W\!P(\boldsymbol{r_h}(t)) - W\!P(\boldsymbol{r_{hit}})], \label{Eq3}\\
		Q_e(t) = -q[W\!P(\boldsymbol{r_e}(t)) - W\!P(\boldsymbol{r_{hit}})], \label{Eq4}
	\end{gather}
	respectively, where $W\!P(\boldsymbol{r_{hit}})$ represents the position at which holes and electrons are generated. As shown in Fig. \ref{fig:6-Figure}(a), the absolute value of $Q_h(t)$ is usually much larger than that of $Q_e(t)$ because $W\!P(\boldsymbol{r_{hit}})$ is below \SI{0.1}{\V} in most areas of the detector, except near the P$^+$ point contact. Therefore, holes make the dominant contribution to the charge signal. According to Eq. \eqref{Eq3}, the time evolution of $Q_h(t)$ depends on the change in $W\!P(\boldsymbol{r_h}(t))$. As the holes drift to the P$^+$ point contact, the change in $W\!P(\boldsymbol{r_h}(t))$ is initially very small but becomes quite large. Thus, $Q_h(t)$ has a slow component followed by a fast component.
	
	Fig. \ref{fig:7-Figure} shows the drift trajectories of an electron-hole pair at different locations inside the detector and their signals. The difference between these signals generated at different locations depends on how long the hole or electron drifts before it is collected by the corresponding electrode. For instance, the hole carriers generated in most areas of the detector must pass through the weak field region in the middle of detector, which causes a long rise time initially. By contrast, the events occur near the passivation layer, and their hole carriers will be collected directly without passing through the weak field region; thus, the rise time is shorter. Therefore, the rising tendency of the waveform can indicate the approximate position of particle interaction. The simulation proves that FBEs originate near the passivation layer, which is consistent with the background spectrum reported above.
	
	\section{Comparison of simulation to experiment}
	
	The CDEX-1B detector was used to further investigate the characteristics of FBEs. A \ce{^{60}Co} source experiment was conducted, and the corresponding simulation was also performed using the Geant4 simulation toolkit \cite{agostinelliGeant4SimulationToolkit2003,allisonRecentDevelopmentsGeant42016} and SAGE package \cite{sheSAGEMonteCarlo2021}.
	
	\subsection{Overview of experiment}
	
\begin{figure}[htbp]
		\centering
		\includegraphics[width=0.9\linewidth]{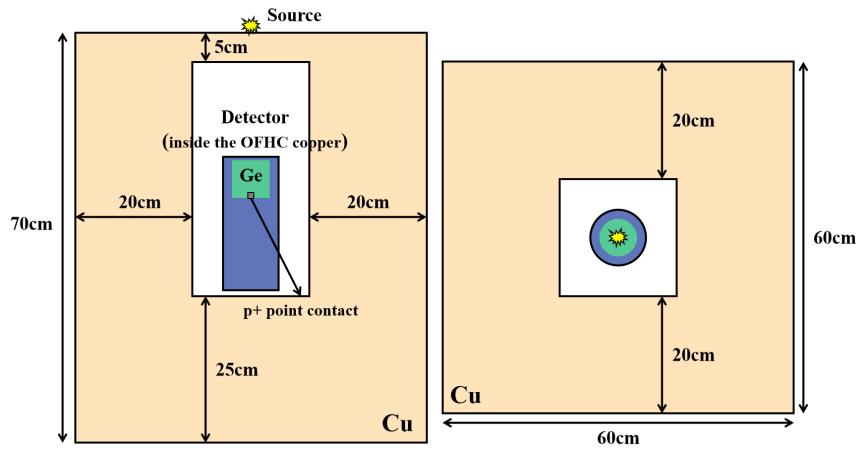}
		\caption{Schematic diagram of CDEX-1B detector and shielding. Left and right panels are side and top views, respectively}
		\label{fig:8-Figure}
	\end{figure}

	The CDEX-1B detector is a \SI{939}{\g} single-element \textit{p}PCGe crystal cylinder with a dead layer of \SI{0.88(12)}{\mm} \cite{maStudyInactiveLayer2017}. The crystal has a diameter of \SI{62.1}{\mm} and a height of \SI{62.3}{\mm}; it is held by a cooper cup with a thickness of \num{2.0} and \SI{1.0}{\mm} on the lateral and top surfaces, respectively. Then the entire structure is mounted inside a copper vacuum cryostat with a thickness of \num{2.0} and \SI{1.1}{\mm} on the lateral and top surface, respectively. The copper bricks acting as shielding outside the detector are \SI{20}{\cm} thick at the sides, \SI{25}{\cm} thick at the bottom, and \SI{5}{\cm} thick at the top, as shown in Fig. \ref{fig:8-Figure}. The shielding has a collimating slit at the top along which $\gamma$ rays can enter.
	
	\begin{figure}[htbp]
		\centering
		\includegraphics[width=1.0\linewidth]{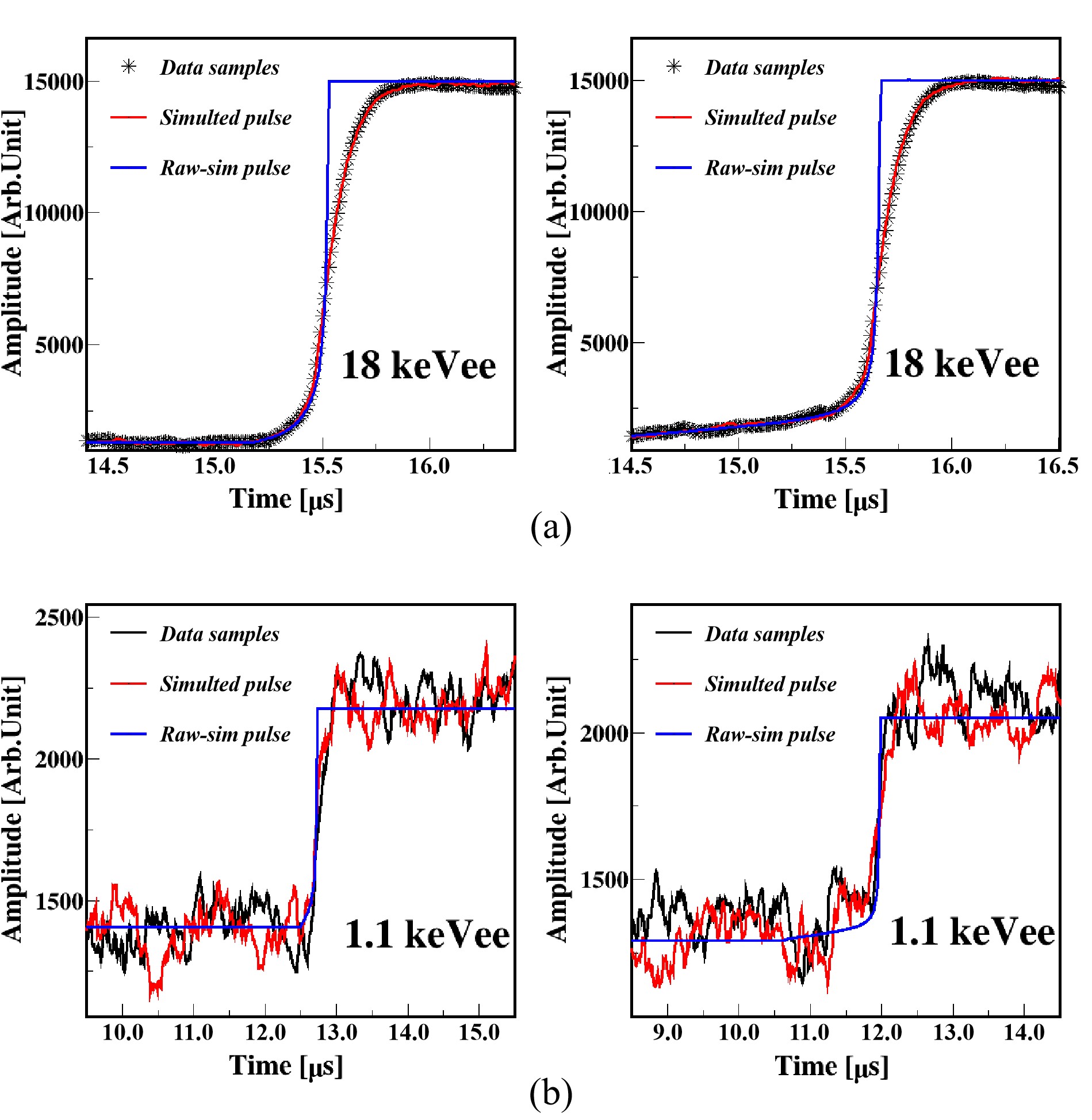}
		\caption{Raw pulses (blue line) and post-processing pulse (red line) with electronics response and electric noise at (a) \SI{18}{\keVee} and (b) \SI{1.1}{\keVee}, with real pulses (black point), where left panel shows an FBE, and right panel shows a BE}
		\label{fig:9-Figure}
	\end{figure}
	
	\begin{figure}[htbp]
		\centering
		\includegraphics[width=1.0\linewidth]{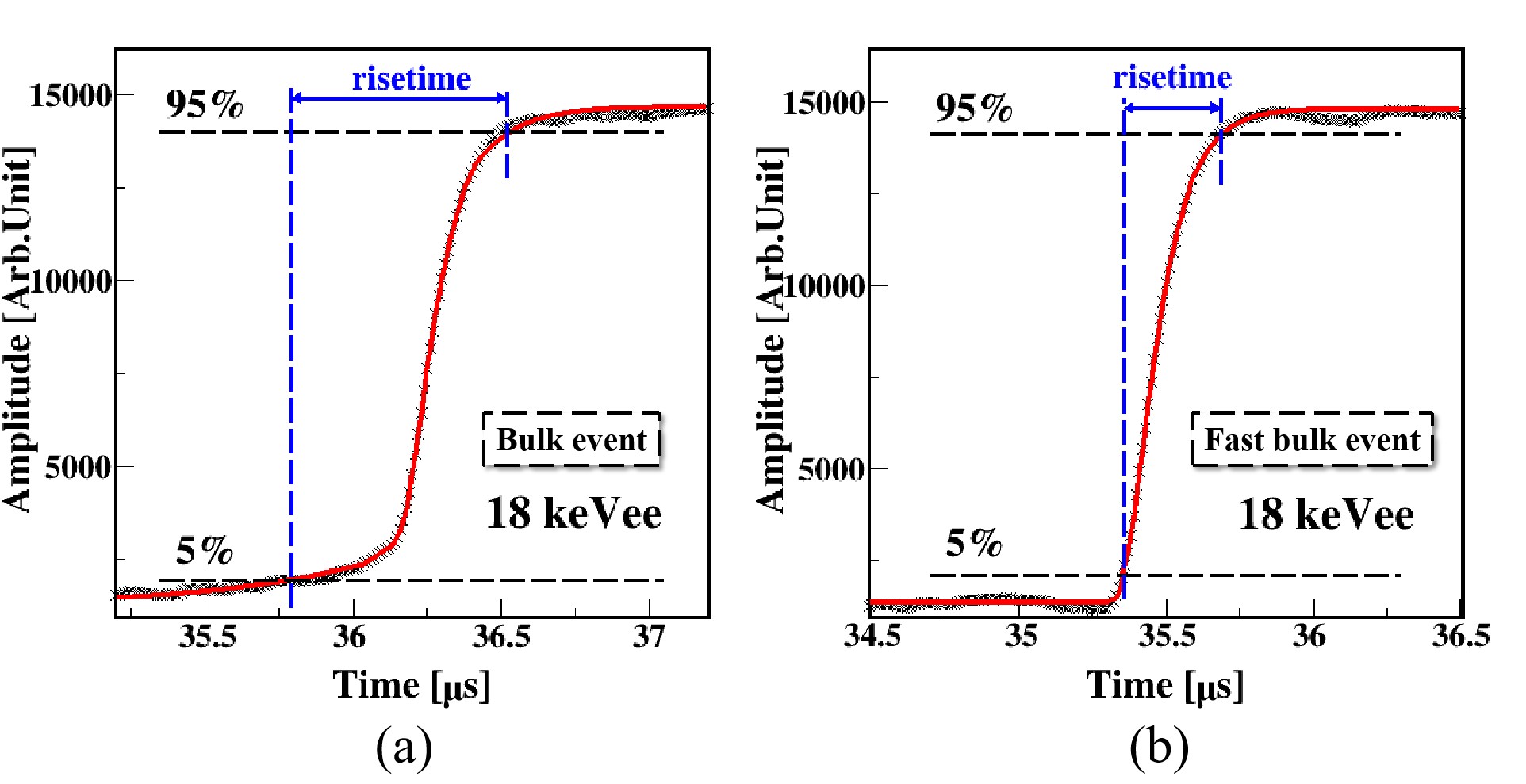}
		\caption{Pulse shape of (a) BE and (b) FBE, where two hyperbolic tangent functions are applied to fit the front and back edges of the waveform, respectively. The rise time $\tau$ is parameterized by the time interval between \SI{5} and \SI{95}{\%} of the amplitude of the fitted function}
		\label{fig:10-Figure}
	\end{figure}
	
	\begin{figure*}[htbp]
		\centering
		\includegraphics[width=0.9\linewidth]{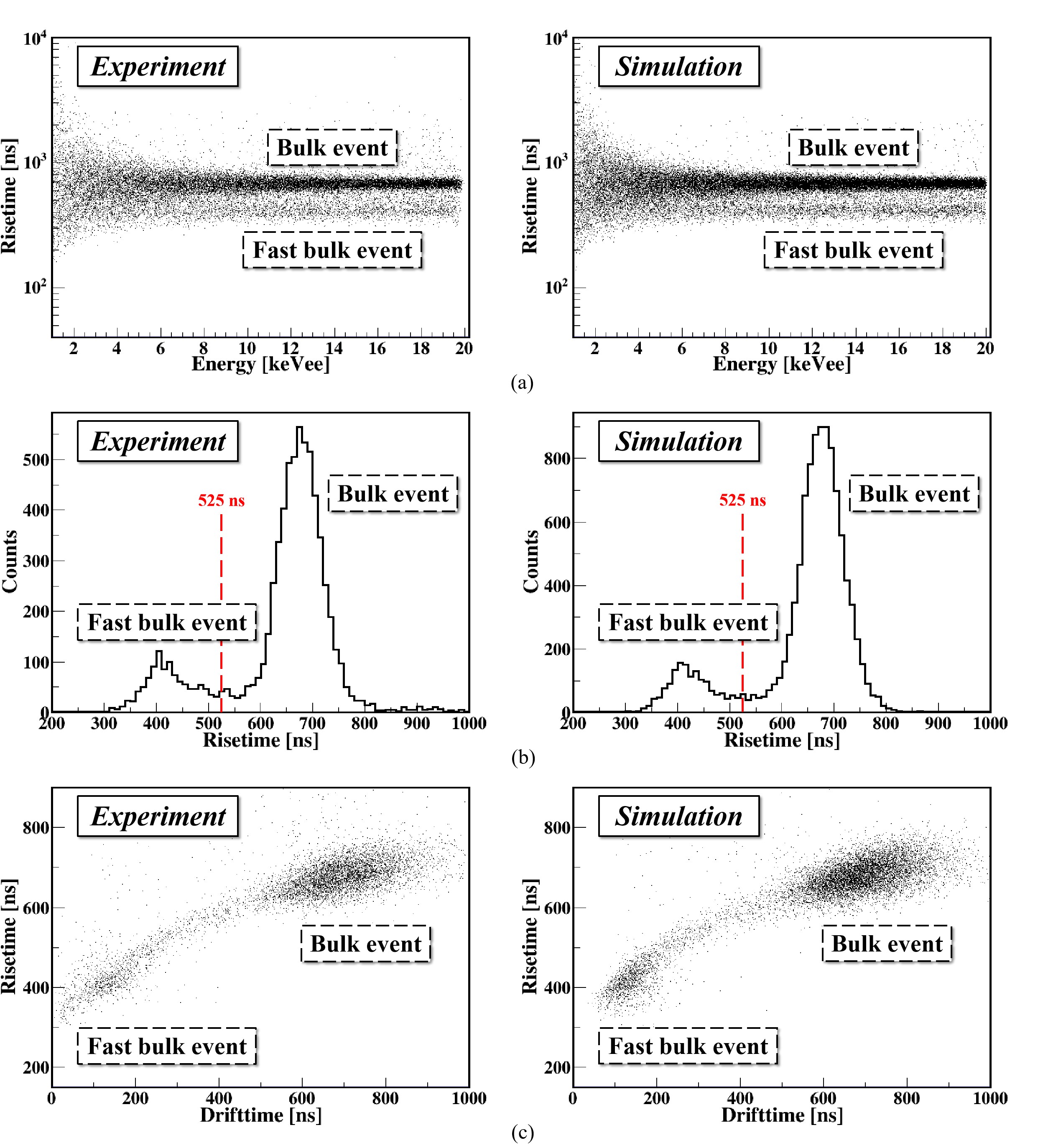}
		\caption{Calibration data (left panels) and simulation results (right panel): (a) scatter plots of rise time versus energy. Two types of events are shown. One, the BE, has a relatively long rise time (\SI{>550}{\ns}). The other is an FBE, with a rise time of approximately \SI{400}{\ns}. (b) Rise time distributions of events with energy above \SI{15}{\keVee}, where the cut at \SI{525}{\ns} is used to distinguish FBEs from BEs. The simulation results are in good agreement with calibration data. (c) Scatter plot of rise time versus drift time for events with energy above \SI{15}{\keVee}. The drift time is parameterized by the time interval between \SI{1} and \SI{5}{\%} of the amplitude of the fitted function. The FBEs, with typical \sisetup{range-phrase = --, range-units = single}\SIrange{300}{500}{\ns} rise time and \SIrange{50}{200}{\ns} drift time, have a faster rising pulse than BEs, which have a typical rise time of \SIrange{600}{800}{\ns} and drift time of \SIrange{500}{800}{\ns}}
		\label{fig:11-Figure}
	\end{figure*}
	
	The P$^+$ point contact is at the bottom of the crystal and is connected to a brass pin to read out the signals \cite{jiangMeasurementDeadLayer2016}. After a pulsed-reset feedback preamplifier, four identical energy-related outputs are generated. Two of them are distributed into two shaping amplifiers (SAs), and the other two are loaded to timing amplifiers (TAs) at different gains. The SAs, which have a high gain, cover the low-energy region (\sisetup{range-phrase = --, range-units = single}\SIrange{0}{12}{\keV}) at \SI{6}{\us} (SA$_6$) and \SI{12}{\us} (SA$_{12}$), respectively. The outputs of the two TAs provide time information, including the high gain in the \SIrange{0}{20}{\keV} region and the low gain in the \SIrange{0}{1.3}{\MeV} region. The data acquisition system is triggered by the output of SA$_6$ or a random trigger generated by a precision pulser \cite{cdexcollaborationFirstResults76Ge2017,AROOTbasedDetector2021}. These signals from the amplifiers are recorded and digitized by a \num{14}-bit \num{100}-\si{\MHz} flash analog-to-digital converter. The RMS of the pedestal is \SI{31}{\eVee}, whereas the full width at half maximum of the test pulser is \SI{80}{\eVee} \cite{yangLimitsLightWIMPs2018}.
	In the measurement, an uncollimated \ce{^{60}Co} source is mounted at the center of the top collimating slit. The average trigger rate is \SI{36.3}{\Hz}, and the run time is $\sim$\SI{3}{\day}. The operational voltage applied to the detector is $+\SI{3750}{\V}$.
	
	\subsection{Simulation procedures}
	
	\begin{figure}[htbp]
		\centering
		\includegraphics[width=0.96\linewidth]{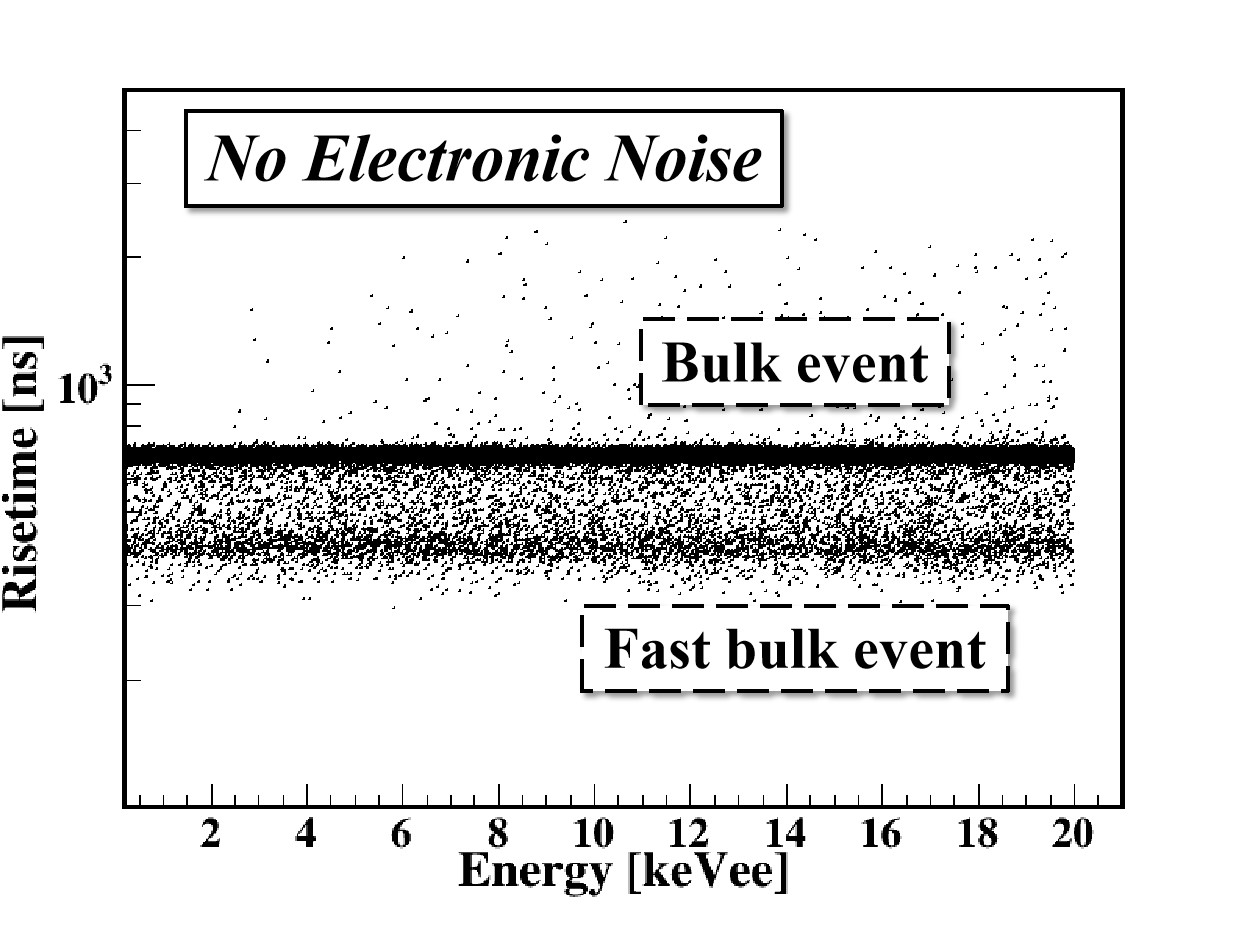}
		\caption{Scatter plot of rise time versus energy in a simulation without electric noise. The result proves that the deposited energy does not depend on the rise time, which depends on the location in the detector}
		\label{fig:12-Figure}
	\end{figure}
	
	In the Geant4 simulation, the realistic positions and deposited energy of $\gamma$ rays emitted by the \ce{^{60}Co} source in the detector were simulated in accordance with the shielding and structures of the CDEX-1B \ce{^{60}Co} experiment mentioned above. 
	
	Using the positions and energies generated by Geant4, an event-by-event pulse shape simulation was conducted using the geometry of the CDEX-1B detector. For each event, the signals corresponding to individual hits were simulated and then summed to obtain the waveform. The N$^+$ electrode of the detector was set to a realistic operating voltage of $+\SI{3750}{\V}$. To obtain simulated pulses that are in agreement with the experimental pulses, the net impurity concentration was set to $-\SI{1.28e10}{\cm}^{-3}$ at the P$^+$ point contact end, and decrease linearly at a rate of $-\SI{0.034e10}{\cm}^{-3}/\si{\cm}$. The pulse shape simulation also considers the nonuniformity of the radial net impurity concentration, as well as changes in charge cloud size due to diffusion and self-repulsion. However, these effects are negligible. For these configurations, the simulated depletion voltage is \SI{3362}{\V}, that is $\sim$\SI{400}{\V} below the real operating voltage, which is a reasonable result.
	
	Finally, post-processing was performed. To simulate real pulse, the electronic response and electrical noise were considered. The electronics response was obtained by feeding the approximate $\delta$ impulse to the test input of the preamplifier. The waveforms of the electrical noise were sampled from random trigger events. The simulation raw pulses and post-processing pulses at two energies are displayed in Fig. \ref{fig:9-Figure} with the real pulses. The simulated and real pulses are in good agreement. The charge collection process for the BE can be separated into slow and fast components. The slow component is determined mainly by the drift time during which the hole remains in the weak electric field. The fast component is determined by the strong filed near the P$^+$ electrode, in which most events exhibit the same behavior. For the FBE, the time required for holes to pass through the weak electric field area is relatively short; consequently, its slow component is too small to be distinguished.
	
	\subsection{Measurement and simulation results} \label{sec.IV-C}
	
	To accurately describe the TA waveform and reduce the noise effect, the pulse is fitted by different hyperbolic tangent functions at the front and back edges, because of the asymmetry of the pulse shape. The fitting results (Fig. \ref{fig:10-Figure}) indicate the fine description of the pulse. The rise time $\tau$ is parameterized by the time interval between \SI{5} and \SI{95}{\%} of the amplitude of the fitted function.
	
	Fig. \ref{fig:11-Figure}(a) shows scatter plots of rise time versus energy for experimental and simulated events. At energies above \SI{6}{\keVee}, the BEs have a relatively long rise time (\SI{>550}{\ns}), and FBEs have a rise time of approximately \SI{400}{\ns}. Below \SI{6}{\keVee}, these two types of events are indistinguishable. Fig. \ref{fig:11-Figure}(b) and (c) shows the rise time distributions and rise time versus drift time scatter plots at energies above \SI{15}{\keVee}, respectively, where the drift time is parameterized by the time interval between \SI{1} and \SI{5}{\%} of the amplitude of the fitted function. All of these results show that the simulation results are in good agreement with calibration data.
	
	\begin{table}[htbp]
		\centering
		\caption{\newline Mean ($\mu$) and width ($\sigma$) of the FBE and BE distributions at different energies}
		\label{tab:1-table}
		\renewcommand\arraystretch{1.5}
		\begin{tabular*}{8.5cm}{@{\extracolsep{\fill}}ccccc}
			\toprule
			\multirow{2}{*}{Energy bin} & \multicolumn{2}{c}{FBE} & \multicolumn{2}{c}{BE} \\
			\cmidrule{2-5}
			& $\mu$ (\si{\ns}) & $\sigma$ (\si{\ns}) & $\mu$ (\si{\ns}) & $\sigma$ (\si{\ns}) \\
			\midrule
			\sisetup{range-phrase = --, range-units = single}\SIrange{18}{19}{\keVee} & \num{409.0} & \num{28.3} & \num{677.6} & \num{37.6} \\
			\SIrange{16}{17}{\keVee} & \num{414.2} & \num{32.7} & \num{675.8} & \num{40.7} \\
			\SIrange{14}{15}{\keVee} & \num{407.6} & \num{39.2} & \num{675.0} & \num{52.5} \\
			\SIrange{10}{11}{\keVee} & \num{412.6} & \num{35.9} & \num{667.7} & \num{69.7} \\
			\SIrange{6}{7}{\keVee}   & \num{437.5} & \num{62.0} & \num{682.0} & \num{97.9} \\
			\SIrange{2}{3}{\keVee}   & \num{521.2} & \num{218.1} & \num{750.8} & \num{295.2} \\
			\SIrange{1}{2}{\keVee}   & \num{581.6} & \num{326.8} & \num{762.7} & \num{395.1} \\
			\bottomrule
		\end{tabular*}
	\end{table}
	
	\begin{figure}[htbp]
		\centering
		\includegraphics[width=0.8\linewidth]{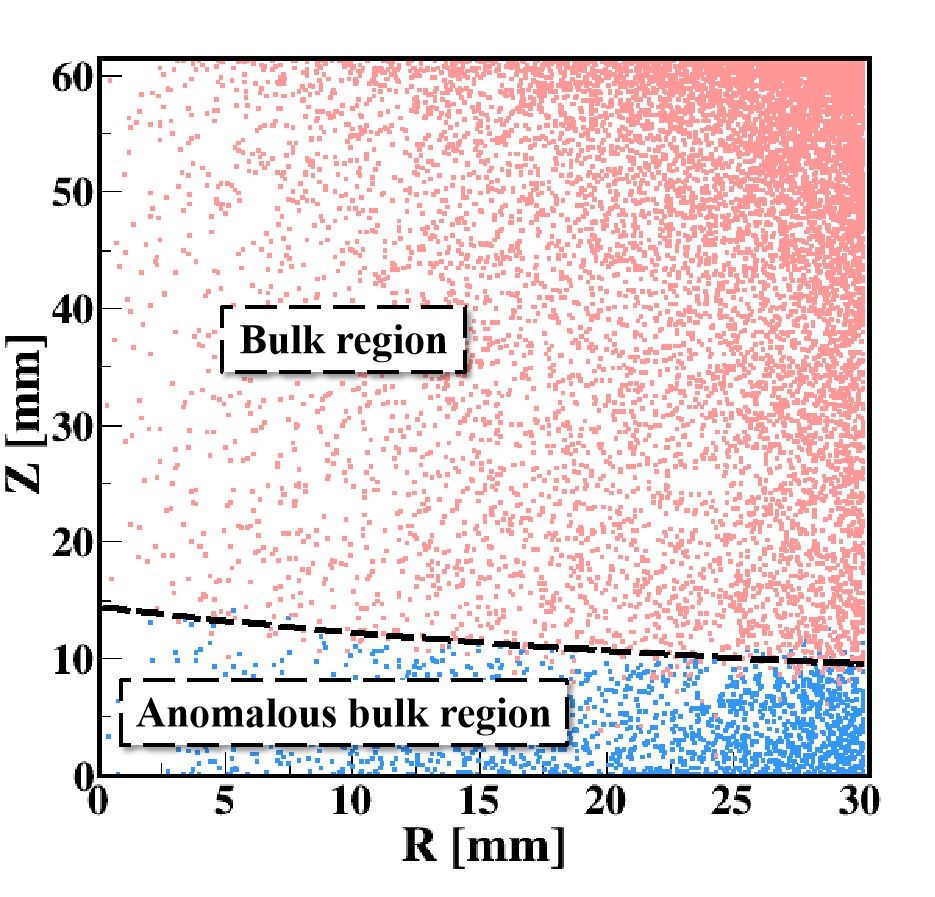}
		\caption{Specific locations of FBEs and BEs in CDEX-1B detector obtained by simulations according to the \SI{525}{\ns} cut above \SI{15}{\keVee}, where the P$^+$ point contact is at $(r = \SI{0}{\mm}, z = \SI{0}{\mm}$). FBEs (blue dots) are distributed near the passivation layer at the bottom of the detector, whereas BEs (red dots) are distributed in most areas of the detector}
		\label{fig:13-Figure}
	\end{figure}
	
	\begin{figure}[htbp]
		\centering
		\includegraphics[width=0.9\linewidth]{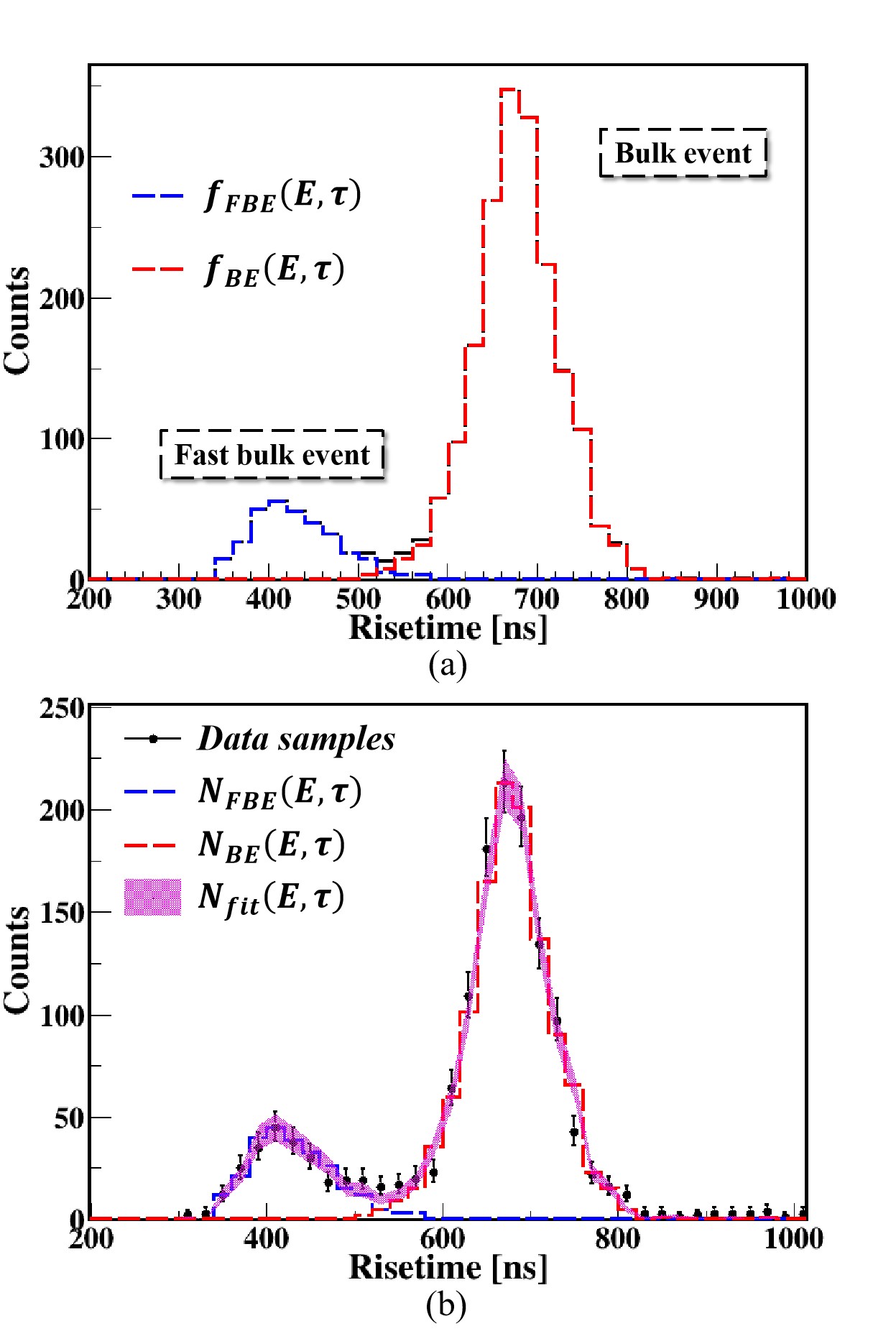}
		\caption{(a) BE (red dotted line) and FBE (blue dotted line) distributions obtained by simulations in the \SIrange{15}{16}{\keVee} range, $f_{B\!E}(E,\tau)$ and $f_{F\!B\!E}(E,\tau)$, respectively. (b) Fitting result in the \SIrange{15}{16}{\keVee} range obtained by fitting the measured $\tau$ distribution $N_{D\!a\!t\!a}(E,\tau)$ (data samples, black circles) with the assumed $\tau$ distribution function $N_{f\!i\!t}(E,\tau)$ (magenta shading). The corresponding $N_{B\!E}(E,\tau)$ (red dotted line) and $N_{F\!B\!E}(E,\tau)$ (blue dotted line) are obtained}
		\label{fig:14-Figure}
	\end{figure}
	
	\begin{figure*}[htbp]
		\centering
		\includegraphics[width=0.95\linewidth]{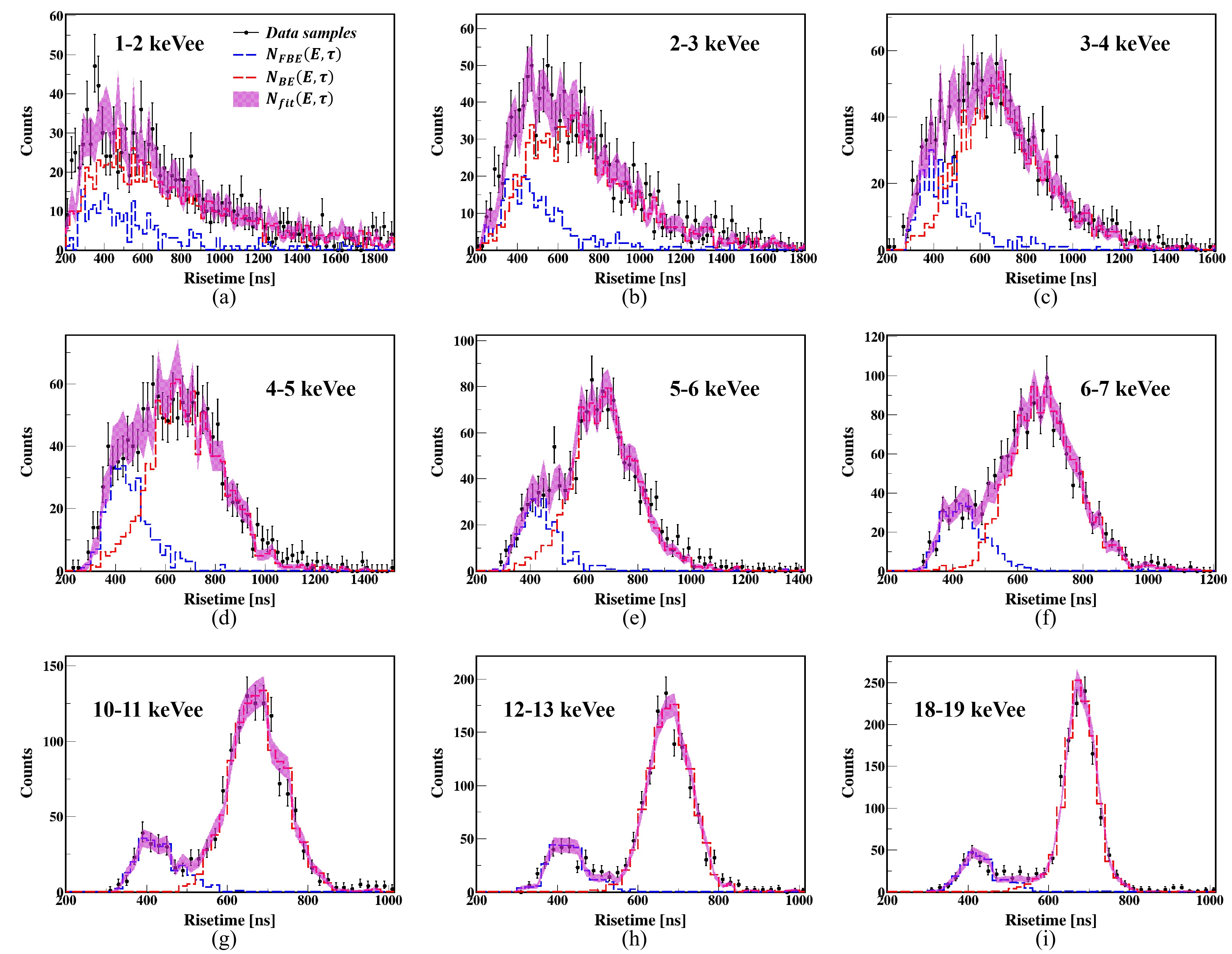}
		\caption{Fitting results in different energy ranges: (a) \SIrange{1}{2}{\keVee}; (b) \SIrange{2}{3}{\keVee}; (c) \SIrange{3}{4}{\keVee}; (d) \SIrange{4}{5}{\keVee}; (e) \SIrange{5}{6}{\keVee}; (f) \SIrange{6}{7}{\keVee}; (g) \SIrange{10}{11}{\keVee}; (h) \SIrange{12}{13}{\keVee}; (i) \SIrange{18}{19}{\keVee}. All results are in good agreement with the experiment}
		\label{fig:15-Figure}
	\end{figure*}
	
	\begin{figure}[htbp]
		\centering
		\includegraphics[width=0.90\linewidth]{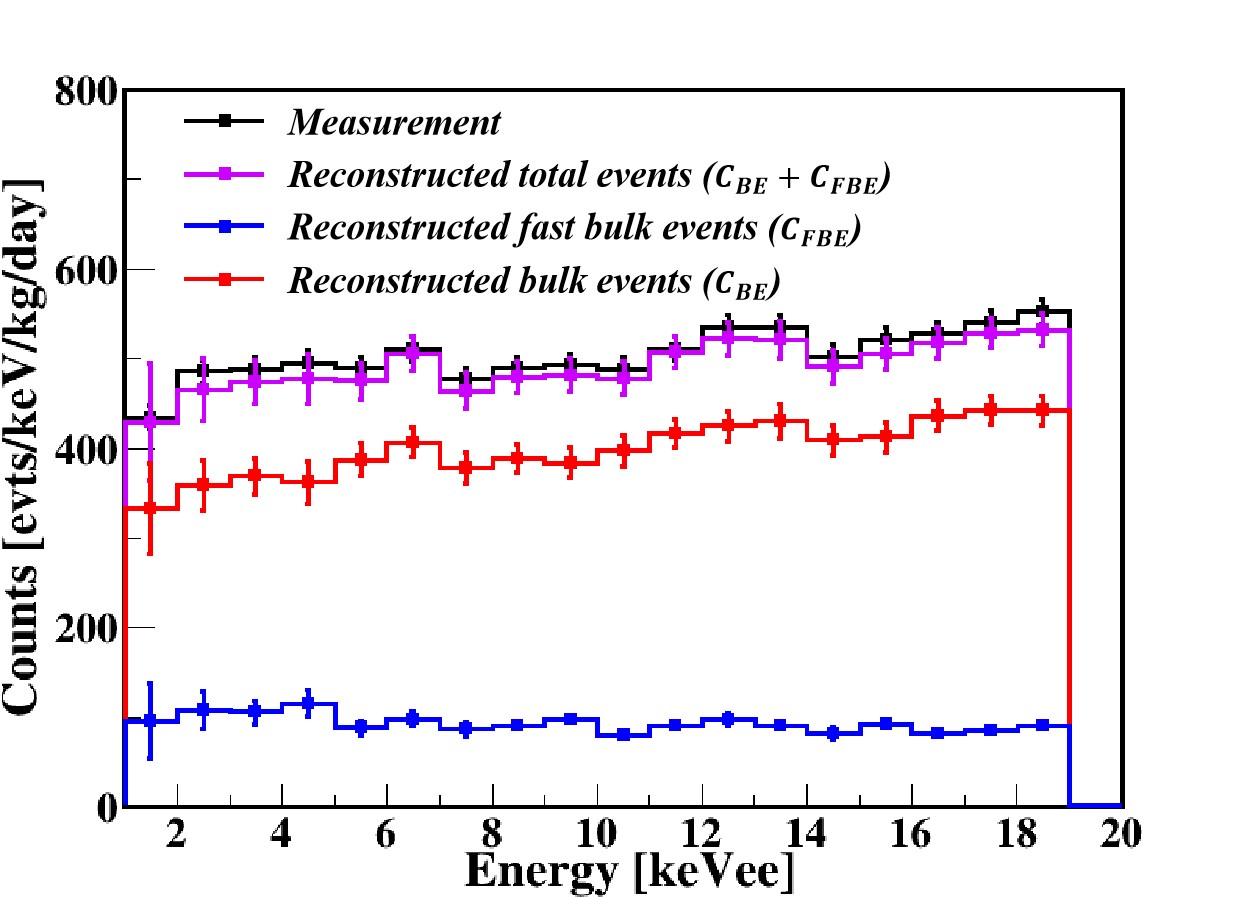}
		\caption{Reconstructed energy spectrum (magenta line), which contains spectral components of BEs (red line) and FBEs (blue line), compared with measured total energy spectrum (black line)}
		\label{fig:16-Figure}
	\end{figure}
	
	\begin{figure}[htbp]
		\centering
		\includegraphics[width=0.90\linewidth]{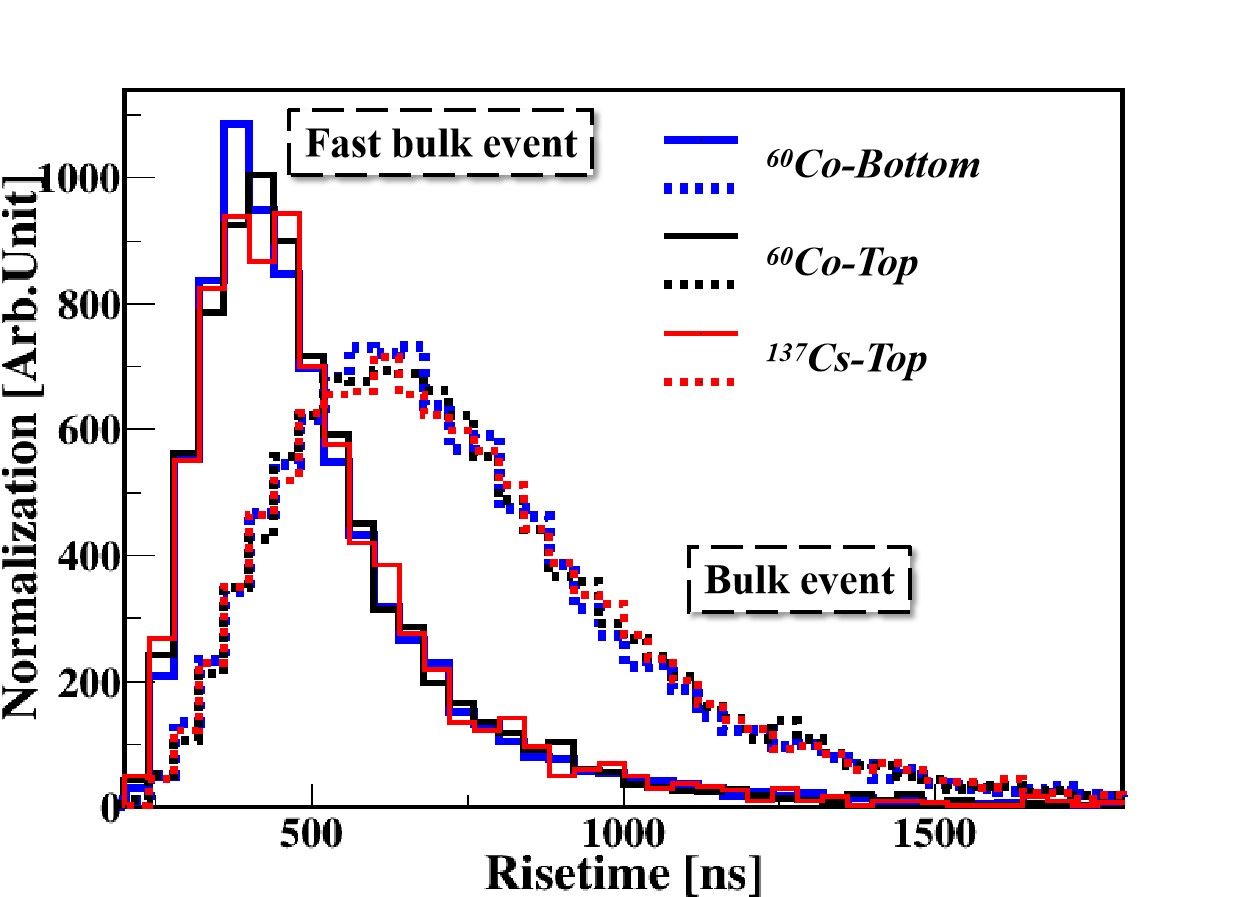}
		\caption{Comparison of $\tau$ distributions at \SIrange{2}{3}{\keVee} of \ce{^{60}Co} source placed at the bottom of the crystal (blue), \ce{^{137}Cs} source located at the top of the collimating slit (red), and \ce{^{60}Co} experiment mentioned in Sec. \ref{sec.IV-C} (black). Their FBE and BE distributions are normalized by the counts of FBEs and BEs, respectively, for a \ce{^{60}Co} source placed at the bottom}
		\label{fig:17-Figure}
	\end{figure}
	
	\begin{figure*}[htbp]
		\centering
		\includegraphics[width=0.9\linewidth]{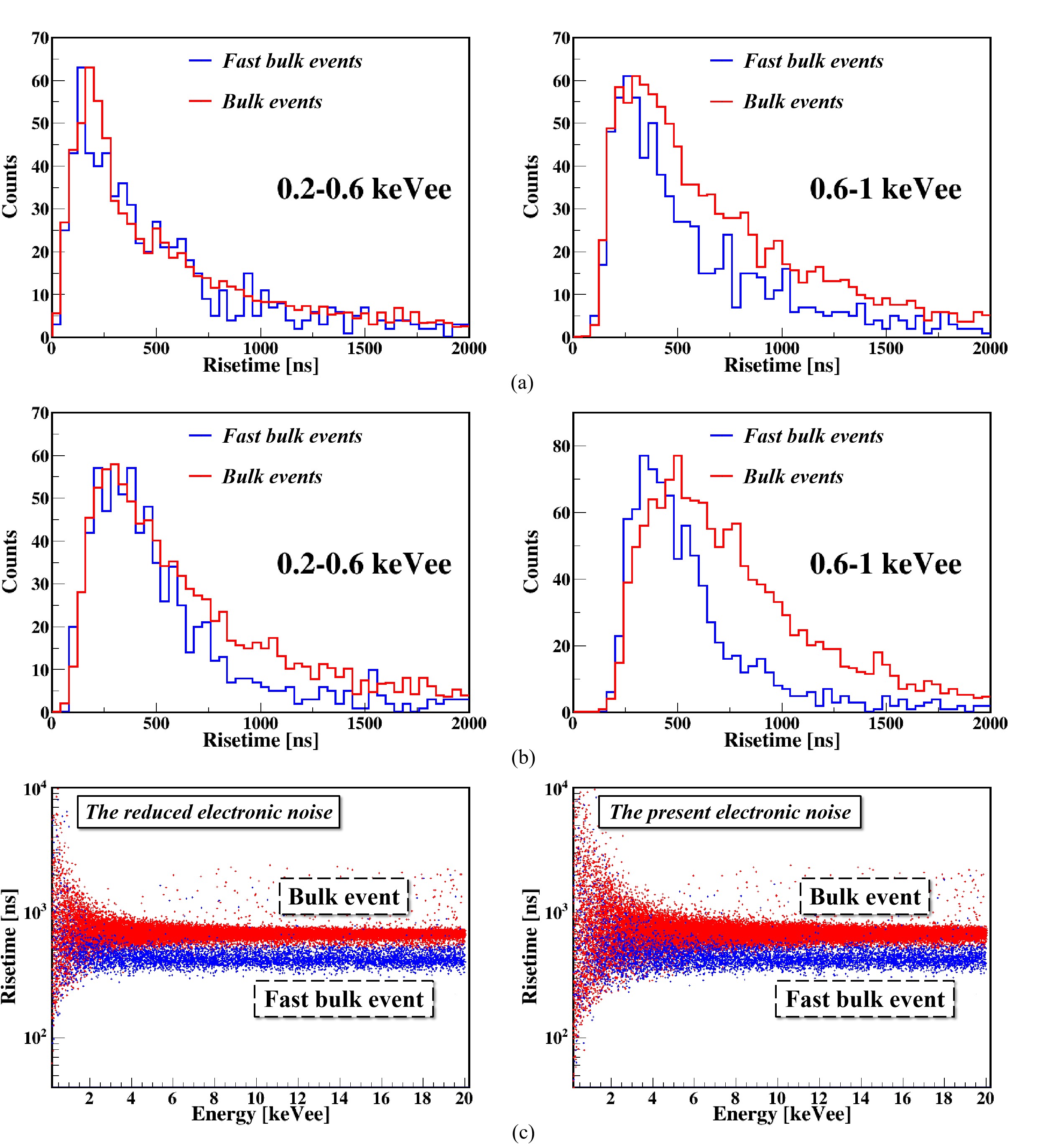}
		\caption{Simulated rise time distributions of FBEs (blue line) and BEs (red line) in the \SIrange{0.2}{0.6}{\keVee} energy range (left panel) and \SIrange{0.6}{1}{\keVee} energy range (right panel) assuming (a) current electronic noise and (b) electronic noise reduced by a factor of two. (c) Scatter plot of rise time versus energy for simulated events with energy above \SI{0.2}{\keVee} assuming an electronic noise level reduced by a factor of two (left panel), where blue points represent FBEs and red points represent BEs. A scatter plots under the current electronic noise (right panel) is presented for comparison}
		\label{fig:18-Figure}
	\end{figure*}
	
	As illustrate in Fig. \ref{fig:11-Figure}(b), the \SI{525}{\ns} cut is used to distinguish FBEs from BEs. The mean ($\mu$) and width ($\sigma$) of the FBE and BE distributions at different energies are presented in Table \ref{tab:1-table}. At relatively high energies, the $\mu$ values of both distributions do not change and are independent of energy. Because of interference caused by electrical noise at energies below \SI{3}{\keVee}, where $\sigma$ is comparable to the band separation, serious cross-contaminations occurs between BEs and FBEs, which causes the bands to merge. However, as shown in Fig. \ref{fig:12-Figure}, the simulated relationship between the rise time and energy without electrical noise proves that the rise time is independent of the deposited energy; consequently, the FBE region is also independent of the deposited energy. Thus, the regions at which FBEs and BEs occur inside the detector can be identified from the simulation result at relatively high energies.
	
	Fig. \ref{fig:13-Figure} shows the simulated locations of these two types of events in the CDEX-1B detector according to the \SI{525}{\ns} cut above \SI{15}{\keVee}. FBEs (blue dots) are distributed near the passivation layer at the bottom of the detector. This result is consistent with the region predicted above. The volume ratio of FBE region to the volume of the entire crystal is \SI{18.8}{\%}, which is consistent with that derived from the \ce{^{68}Ge} peak \sisetup{separate-uncertainty}\SI{16.2(30)}{\%} and \ce{^{65}Zn} peak \SI{19.4(52)}{\%}. Both isotopes are cosmogenic nuclides and are expected to be uniformly distributed in germanium crystals.
	
	\section{BE and FBE discrimination}
	
	\subsection{Formulation}
	
	The goal of this analysis is to discriminate BEs from FBEs and extract their energy spectra in the \ce{^{60}Co} experiment. The reconstructed count rates $C_{B\!E}(E)$ and $C_{F\!B\!E}(E)$ are given by
	\begin{equation}
		\begin{aligned}
			C_{B\!E}(E) & = \int_{all\,\tau}N_{B\!E}(E,\tau)\,\mathrm{d}\tau, \\
			C_{F\!B\!E}(E) & = \int_{all\,\tau}N_{F\!B\!E}(E,\tau)\,\mathrm{d}\tau.
		\end{aligned}
	\end{equation}
	In particular, $C_{B\!E}(E)$ is the WIMP- and neutrino-induced candidate spectrum when background events are analyzed. $N_{B\!E}(E,\tau)$ and $N_{F\!B\!E}(E,\tau)$ represent the reconstructed count rates of BEs and FBEs, respectively, at energy $E$ and rise time $\tau$. They are defined as follows:
	\begin{equation}
		\begin{aligned}
			N_{f\!i\!t}(E,\tau) & = N_{B\!E}(E,\tau) + N_{F\!B\!E}(E,\tau) \\
			& = p_0(E) \cdot f_{B\!E}(E,\tau) + p_1(E) \cdot f_{F\!B\!E}(E,\tau),
		\end{aligned}
	\end{equation}
	where $f_{B\!E}(E,\tau)$ and $f_{F\!B\!E}(E,\tau)$ are the unnormalized probability density functions (PDFs) of the rise time distributions of BEs and FBEs obtained from simulations using the same settings as the \ce{^{60}Co} experiment mentioned above. In addition, $p_0(E)$ and $p_1(E)$ are $\tau$-independent scaling factors, which are proportional to the corresponding count rate ratios. These two parameters are obtained by fitting the measured $\tau$ distribution $N_{D\!a\!t\!a}(E,\tau)$ with an additional unitary constrain of $\int_{\tau_1}^{\tau_2}N_{f\!i\!t}(E,\tau)\,\mathrm{d}\tau = \int_{\tau_1}^{\tau_2}N_{D\!a\!t\!a}(E,\tau)\,\mathrm{d}\tau$ in the fitting interval $(\tau_1,\tau_2)$.
	
	\subsection{Fitting results}
	
	\begin{table*}[htbp]
		\centering
		\caption{\newline Contributions of various sources to statistical and systematic uncertainties of FBE energy spectrum in \SIrange{1}{2}{\keVee} and \SIrange{2}{3}{\keVee} energy bins}
		\label{tab:2-table}
		\renewcommand\arraystretch{1.5}
		\begin{tabular*}{17.8cm}{@{\extracolsep{\fill}}ccc}
			\toprule
			Energy bin & \SIrange{1}{2}{\keVee} & \SIrange{2}{3}{\keVee} \\
			\midrule
			\multicolumn{1}{l}{Counts and total error (\si{\per\kg\per\keV\per\day})} & \num{96.16(4122)} & \num{107.68(2067)} \\
			\multicolumn{1}{l}{I) Statistical uncertainties} & & \\
			\multicolumn{1}{l}{\qquad\quad $p_0(E)$ and $p_1(E)$} & \num{26.55} & \num{15.92} \\
			\multicolumn{1}{l}{II) Systematic uncertainties} & & \\
			\multicolumn{1}{l}{\qquad\quad Choice of $f_{B\!E}(E,\tau)$ and $f_{F\!B\!E}(E,\tau)$} & \num{21.73} & \num{8.64} \\
			\multicolumn{1}{l}{\qquad\quad Choice of fitting interval $(\tau_1,\tau_2)$} & \num{14.37} & \num{2.36} \\
			\multicolumn{1}{l}{\qquad\quad Size of $\tau$ bin} & \num{17.76} & \num{9.67} \\
			Combined systematic uncertainties & \num{31.53} & \num{13.18} \\
			\bottomrule
		\end{tabular*}
	\end{table*}
	
	\begin{table*}[htbp]
		\centering
		\caption{\newline Mean ($\mu$) and width ($\sigma$) of FBE and BE distributions in \SIrange{0.2}{0.6}{\keVee} and \SIrange{0.6}{1}{\keVee} energy bins for current electronic noise and electronic noise reduced by a factor of two}
		\label{tab:3-table}
		\renewcommand\arraystretch{1.5}
		\begin{tabular*}{17.8cm}{@{\extracolsep{\fill}}c*{8}{p{1.78cm}}}
			\toprule
			\multirow{3}{*}{Energy bin} & \multicolumn{4}{c}{Current electronic noise} & \multicolumn{4}{c}{Reduced electronic noise} \\
			\cmidrule{2-9}
			& \multicolumn{2}{c}{FBE} & \multicolumn{2}{c}{BE} & \multicolumn{2}{c}{FBE} & \multicolumn{2}{c}{BE} \\
			\cmidrule{2-9}
			& \multicolumn{1}{c}{$\mu$ (\si{\ns})} & \multicolumn{1}{c}{$\sigma$ (\si{\ns})} & \multicolumn{1}{c}{$\mu$ (\si{\ns})} & \multicolumn{1}{c}{$\sigma$ (\si{\ns})} & \multicolumn{1}{c}{$\mu$ (\si{\ns})} & \multicolumn{1}{c}{$\sigma$ (\si{\ns})} & \multicolumn{1}{c}{$\mu$ (\si{\ns})} & \multicolumn{1}{c}{$\sigma$ (\si{\ns})} \\
			\midrule
			\SIrange{0.2}{0.6}{\keVee} & \multicolumn{1}{c}{\num{524.7}} & \multicolumn{1}{c}{\num{440.8}} & \multicolumn{1}{c}{\num{563.4}} & \multicolumn{1}{c}{\num{472.7}} & \multicolumn{1}{c}{\num{531.7}} & \multicolumn{1}{c}{\num{379.0}} & \multicolumn{1}{c}{\num{660.6}} & \multicolumn{1}{c}{\num{446.9}} \\
			\SIrange{0.6}{1}{\keVee} & \multicolumn{1}{c}{\num{582.3}} & \multicolumn{1}{c}{\num{407.5}} & \multicolumn{1}{c}{\num{681.6}} & \multicolumn{1}{c}{\num{440.6}} & \multicolumn{1}{c}{\num{565.8}} & \multicolumn{1}{c}{\num{322.3}} & \multicolumn{1}{c}{\num{768.7}} & \multicolumn{1}{c}{\num{394.9}} \\
			\bottomrule
		\end{tabular*}
	\end{table*}
	
	Fig. \ref{fig:14-Figure} shows the best-fit result for the rise time distribution in the CDEX-1B \ce{^{60}Co} experiment at \sisetup{range-phrase = --, range-units = single}\SIrange{15}{16}{\keVee} and the PDFs [$f_{B\!E}(E,\tau)$, $f_{F\!B\!E}(E,\tau)$] obtained from the simulation. The fitted $\tau$ distribution $N_{f\!i\!t}(E,\tau)$ is in good agreement with that of the calibration data, with a total $\chi^2/n_d$ of $29.8/24$ (where $n_d$ represents the degrees of freedom). The results in Fig. \ref{fig:15-Figure} show good agreement with the experiment, even in the low-energy region below \SI{4}{\keVee}.
	
	The energy spectra of BEs and FBEs were reconstructed on the basis of these results, as shown in Fig. \ref{fig:16-Figure}. The reconstructed and measured BE energy spectra show a consistent trend. In the low-energy region (\SI{<5}{\keVee}), the FBE spectrum may exhibit a slight increase, which may result from SEs in the passivation layer. In addition, the reconstructed event count rate is slightly lower than the calibration data, because Compton events in which one hit occurs in the dead layer and the other in the bulk/fast bulk region are not counted. These events exhibit a normal drift time, like that of BEs/FBEs, and an extremely long rise time at the rear of the pulse, which results in a rise time $\tau$ that is typical of the region above \SI{1000}{\ns}.
	
	Table \ref{tab:2-table} summarizes the statistical and systematic uncertainties of the FBE energy spectrum in the \SIrange{1}{2}{\keVee} and \SIrange{2}{3}{\keVee} energy bins. The total uncertainties of all energy bins in Fig. \ref{fig:16-Figure} are given by combining these uncertainties and applying error propagation.
	
	The fitting parameters $p_0(E)$ and $p_1(E)$ contribute to the statistical uncertainties, which are assigned according to the minimum $\chi^2$ based on the fitting of the experiment and simulation.
	
	The systematic uncertainties are derived as follow.
	
	\begin{enumerate}
		\item[(1)] \textbf{Choice of $f_{B\!E}(E,\tau)$ and $f_{F\!B\!E}(E,\tau)$.} Statistical fluctuations are present in the simulated PDFs, $f_{B\!E}(E,\tau)$, and $f_{F\!B\!E}(E,\tau)$. The systematic uncertainties are given by the deviations of multiple simulation results. They make the main contribution to the systematic uncertainties below \SI{4}{\keVee}, where the statistical fluctuations of PDFs significantly affect the fitting result because of severe merging of FBEs and BEs.
		\item[(2)] \textbf{Choice of fitting interval $(\tau_1,\tau_2)$.} The fitting interval is selected on the basis of sufficient statistics of the $\tau$ distributions in both the experiment and the simulation. The $\tau_2$ values in Table \ref{tab:1-table} are from the BE distribution between ($\mu + \sigma$) to ($\mu + 4\sigma$).
		\item[(3)] \textbf{Size of $\tau$ bin.} The deviations of the results caused by the variation in the size of the $\tau$ bin are regarded as systematic uncertainties. The size of the $\tau$ bin ranges from half to twice the nominal size.
	\end{enumerate}
	
	\subsection{Discussion}
	
	We observe that the \textit{p}PCGe is capable of a single-hit spatial resolution below \SI{20}{\keVee} and can roughly classify events as FBEs and BEs. However, as the energy decreases, the resolution is degraded by noise. Fig. \ref{fig:17-Figure} shows the $\tau$ distributions at \SIrange{2}{3}{\keVee} of the \ce{^{60}Co} source placed at the bottom of the crystal and the \ce{^{137}Cs} source located at the top of the collimating slit, which are compared with that in the \ce{^{60}Co} experiment mentioned in Sec. \ref{sec.IV-C}. The results demonstrate that because of poor spatial resolution, the $\tau$ distributions of both FBEs and BEs are almost independent of the source energy and location. Therefore, when FBEs and BEs below \SI{5}{\keVee} are distinguished, the choice of PDF [$f_{B\!E}$ and $f_{F\!B\!E}$] is universal; that is, it is not necessary to specify the type and position of the source. This fingding is helpful for identifying FBEs and BEs in background events in the low-energy region, which have various unknown and complex sources.
	
	To study the distinction between FBEs and BEs below \SI{1}{\keVee}, where the elimination of SEs is not involved in this work, we present the simulated expectations for their rise time distributions, as shown in Fig. \ref{fig:18-Figure}. In addition, Table \ref{tab:3-table} also lists the mean ($\mu$) and width ($\sigma$) of the FBE and BE distributions for the two energy regions. Both results show subtle and insignificant differences between them, especially below \SI{0.6}{\keVee}. These differences make it challenging to differentiate BEs from FBEs and SEs near the threshold or under conditions of SE penetration and low background statistics.
	
	For comparison, a simulation assuming that the electronic noise is reduced by a factor of two proves that better discrimination can be achieved, as shown by the scatter plot above \SI{0.2}{\keVee} in Fig. \ref{fig:18-Figure}(c), where the scatter plot under the current electronic noise is also depicted. The $\mu$ and $\sigma$ values of the corresponding rise time distributions at \SIrange{0.2}{1}{\keVee} energy are presented in Table \ref{tab:3-table}, and the rise time distributions below \SI{1}{\keVee} are shown in Fig. \ref{fig:18-Figure}(b). In addition, the simulation shows that the current electronic responses of CDEX-1B and CDEX-10 are sufficiently rapid; thus, further response improvement will not significantly increase the ability to distinguish BEs and FBEs. Therefore, reducing the noise of the electronic system is expected to further improve the discrimination of FBEs and BEs at low energies.
	
	\section{Summary and prospect}
	
	FBEs with extremely short rise times were observed in several CDEX experiments. A thorough study was conducted to better understand the wave formation, characteristics, and location distribution of FBEs using a semiconductor electric field and pulse shape simulation. The simulation results were compared with those of a source calibration experiment. FBEs in the CDEX-1B detector occurred near the passivation layer in a region with a volume ratio of \SI{18.8}{\%}, which is consistent with that derived from the \ce{^{68}Ge} peak \SI{16.2(30)}{\%} and \ce{^{65}Zn} peak \SI{19.4(52)}{\%}. Both of these isotopes are expected to be uniformly distributed in germanium crystals. 
	
	This study demonstrates an important property of \textit{p}PCGe, namely, single-hit bulk spatial resolution. In the low-energy region, this resolution makes it possible to roughly discriminate FBEs and BEs, and we provide a method of eliminating FBEs from BEs. These events occur close to the materials around the P$^+$ electrode, which would be among the main background sources in dark matter experiments. Below \SI{1}{\keVee}, extraction of FBEs is challenging under the conditions of SE penetration and low background statistics, and further research on preamplifier noise reduction, data analysis methods, and the optimization of the parameters for extracting FBEs is underway. In the high-energy region, better spatial resolution is expected because of the relatively small effect of electrical noise. Improved resolution is expected to be helpful for identifying the origins of the background and designing new detectors for deployment in CDEX-50dm. Therefore, research on the high-energy spatial resolution of single-hit source experiments and simulations is being pursued. A novel technique based on pulse shape comparison scanning (PSCS) \cite{crespiNovelTechniqueCharacterization2008} will provide a guide for defining the spatial resolution capability of a \textit{p}PCGe detector by scanning the entire detector with a collimated \ce{^{137}Cs} gamma source. A feasibility analysis of the application of PSCS to the CDEX-1B detector has been conducted using simulations, and the relevant experiments are being prepared.
	
	The simulations and measurements presented here improve our understanding of the anomalous bulk effect in \textit{p}PCGe detectors. They serve as a basis for better identification of FBEs as background in rare-event searches with germanium detectors.

\end{document}